\documentclass{kapproc} 

\usepackage{procps} 

\usepackage[dvips]{graphicx}

\upperandlowercase
\kluwerbib 

\usepackage{amsmath}
\usepackage{amssymb}

\begin{document}

\articletitle[Evolution of disk galaxies]
{The evolution of disk galaxies in clusters and the field}

\author{Steven P. Bamford,\altaffilmark{1}$^{*}$
Alfonso Arag\'on-Salamanca,\altaffilmark{1}
and Bo Milvang-Jensen\altaffilmark{2}}

\altaffiltext{1}{School of Physics and Astronomy,
University of Nottingham,\\
University Park,
Nottingham,
NG7 2RD,
UK}

\altaffiltext{2}{
Max-Planck-Institut f\"ur extraterrestrische Physik,\\
        Giessenbachstra\ss e, 85748 Garching, Germany}

\email{$^{*}$ppxspb@nottingham.ac.uk}
\begin{abstract}
  We describe our project to examine the evolution of distant disk galaxies,
  and present the results of our work based on the
  Tully-Fisher relation.  Comparing matched cluster and field samples we find
  evidence that the cluster galaxies are on average $0.7\pm0.2$ mag brighter
  than those in the field.  Considering the field sample alone we find a
  brightening with redshift amounting to $1.0\pm0.5$ mag by $z=1$, which is
  likely to be an upper limit, considering the selection effects.  
  We also give brief details of the ESO Distant Clusters Survey (EDisCS) and
  describe our plans for expanding our studies using data from this survey.
\end{abstract}


\section{Our project}
The broad aim of this project is to investigate disk galaxy evolution, as a
function of both redshift and environment, using rotation velocity as a
baseline for the comparison of various galaxy properties.  To achieve this we
have obtained VLT/FORS2 multi-slit spectroscopy for a large number of bright
emission-line galaxies in the fields of six clusters with redshifts $0.2
\lesssim z_{cl} \lesssim 0.8$.  Rotation velocities were measured by
fitting the [OII] emission line using a synthetic rotation curve method called
\textsc{ELFIT2PY}, which is based on the algorithms of \textsc{ELFIT2D} by
Simard \& Pritchet (1999).  The results of these fits have been thoroughly
examined to determine their quality, and to reject measurements which are
deemed unreliable.  Photometry and structural parameters are measured on
imaging from a range of sources, primarily FORS2 and HST.  Full details of the
sample selection and data analysis are given in Bamford et al. (2005a) and
Milvang-Jensen et al. (2003).  Our final sample consists of 111 emission-line
galaxies with secure rotation velocity measurements, and covers the redshift
range $0.1 \lesssim z \lesssim 1$.

\section{Cluster study}
There is now much evidence indicating a transformation of star-forming spirals
into passive S0 galaxies in cluster environments. However, the detailed form
of this evolution and the mechanisms responsible are still far from certain.
An interesting possibility is that the interaction causes an enhancement of
the star-formation rate prior to its eventual truncation. To examine this
question, we have compared the B-band Tully-Fisher
($M_B\,$--$\,\log{V_\textrm{rot}}$) relation (TFR) for `matched' samples of
field and cluster spirals (\cite{BamfordCluster}).

Fig.~1 shows our TFR for 58 field and 22 cluster galaxies covering similar
ranges in redshift ($0.25 \leq z \leq 1.0$) and luminosity ($M_B \leq -19.5$
mag).  The cluster galaxies show an offset from the field sample, such that
they are $0.7 \pm 0.2$ mag brighter for a given rotation velocity.  Fig.~2
presents this in terms of residuals from the local fiducial relation of Pierce
\& Tully (1992; hereafter PT92), versus redshift.  The offset persists,
at a $3\sigma$ significance level, even when evolution in the field sample is
accounted for, as shown by Fig.~3.

This result implies a phase of enhanced luminosity (and hence star-formation)
in the early stages of a spiral's interaction with the cluster environment.

\begin{figure}
\begin{minipage}{\textwidth}%
\begin{minipage}[t]{0.45\textwidth}%
\includegraphics[width=\textwidth,angle=270]{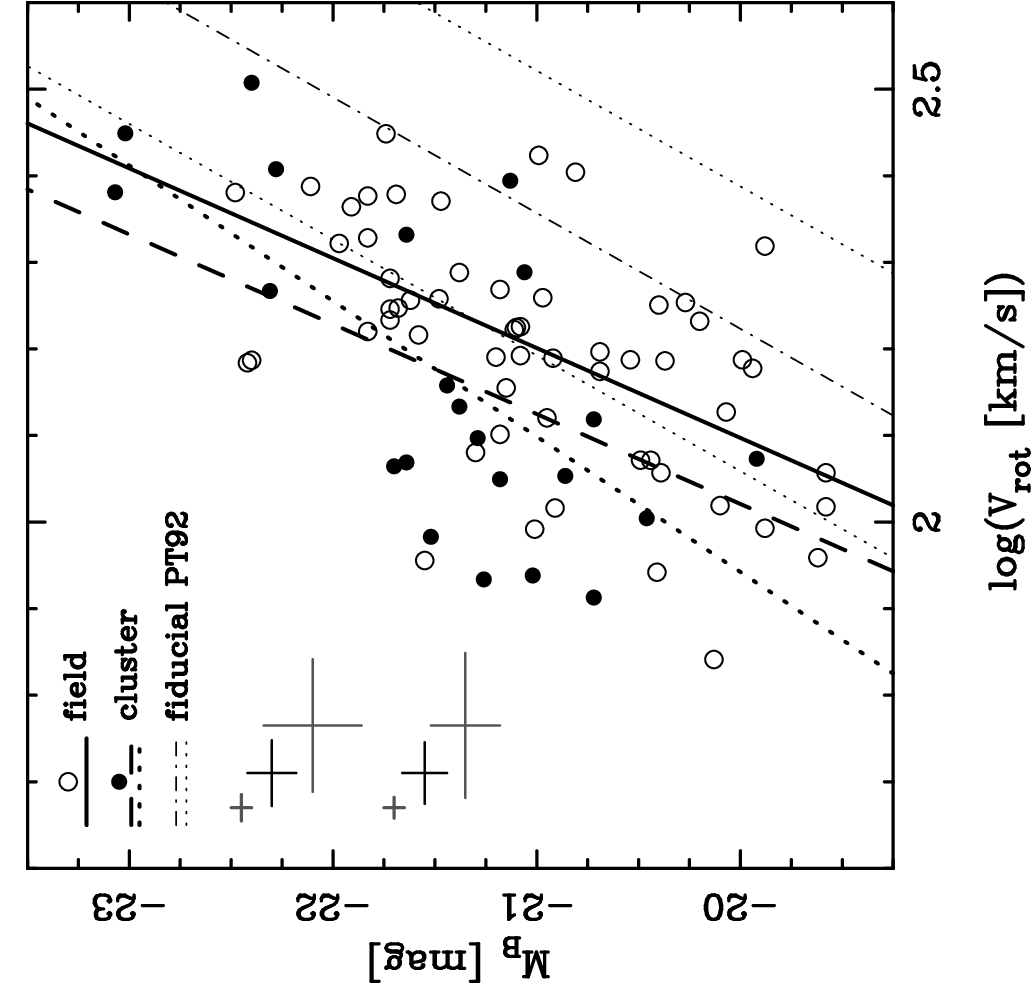}%
\caption{%
  The TFR for our matched samples of field (open points) and cluster (filled
  points) galaxies.  The thin dot-dashed line indicates the fiducial local
  relation of PT92, with its $3\sigma$ scatter delimited by thin dotted lines.
  Fits to the matched field sample (solid line) and cluster sample
  (constrained to the field slope: dashed line, free slope: dotted line) are
  marked.  The two sets of error bars indicate the \mbox{10th-,} 50th- and
  90th-percentile errors for field (top) and cluster (bottom) points.}
\end{minipage}
\hfill%
\begin{minipage}[t]{0.45\textwidth}%
\includegraphics[width=\textwidth,angle=270]{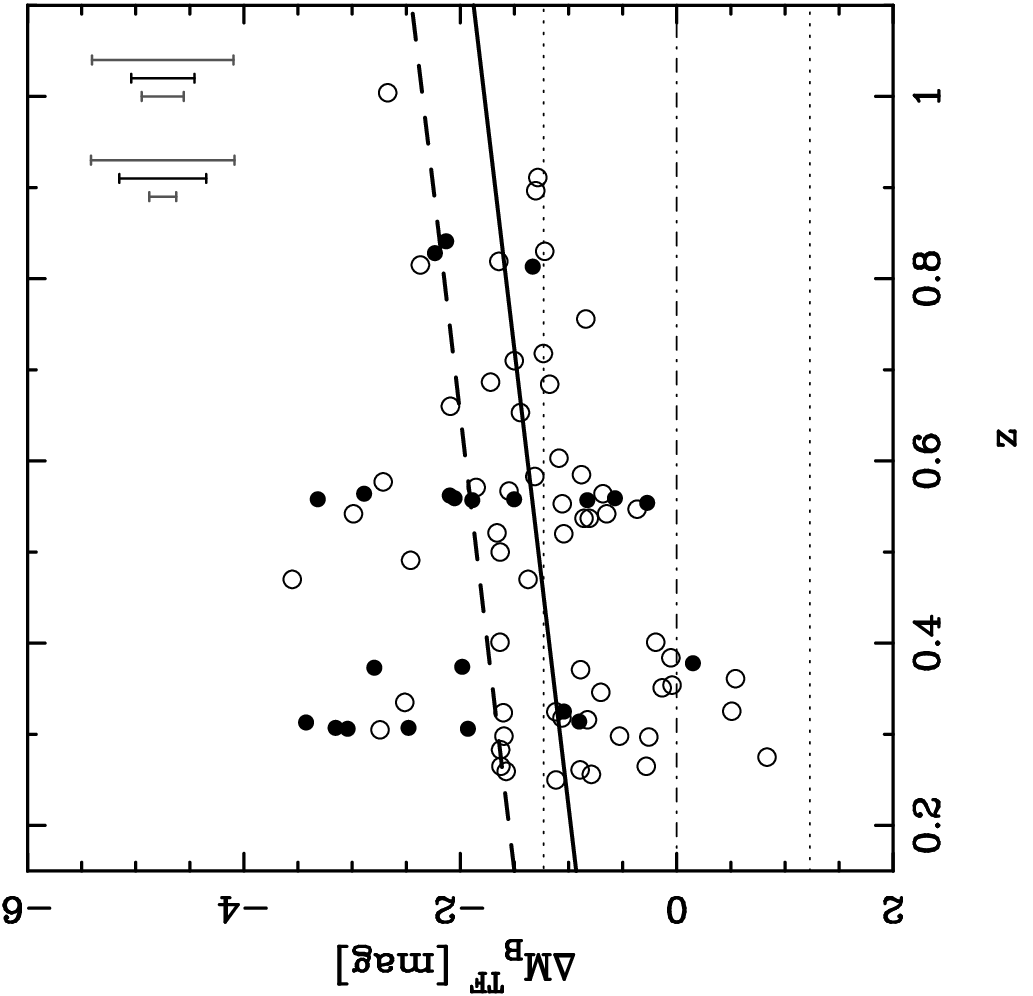}%
\caption{%
  The residuals from the fiducial TFR of PT92 for our
  matched TFR samples of field (open points) and cluster (filled points)
  galaxies.  The thin lines correspond to those in Fig.~1.  Thick
  lines show fits to the full field TFR sample (solid line) and the cluster
  sample, constrained to the field slope (dashed line).  The two sets of error
  bars indicate the 10th-, 50th- and 90th-percentile errors for field (left)
  and cluster (right) points.}
\end{minipage}
\end{minipage}
\end{figure}

\begin{figure}
  \centering
  \includegraphics[width=0.35\textwidth,angle=270]{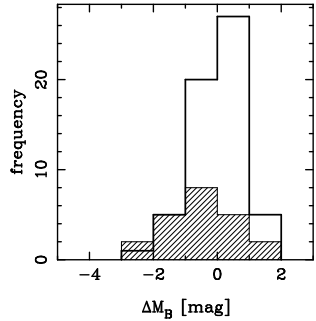}
\caption{Histogram of the residuals of our field (open histogram) and cluster 
  (hatched histogram) data from the fiducial TFR of PT92, with field evolution
  subtracted (i.e., the residuals from the solid line in Fig.~2).
  Note that the cluster distribution is clearly offset from that of the field.}
\end{figure}

\section{Field study}
Studies examining the evolution of the global star formation rate (SFR)
density of the universe find a substantial increase of a factor of $\geq 10$
from $z=0$--$1$ (e.g., \cite{H04}).  Is this trend simply due to an evolution
in the star-formation rates of individual giant spirals, or is some other
process or population responsible?  Does the evolution depend on galaxy mass?

To investigate this we have again used the TFR, to look for changes in the
relation between $M_B$ (a proxy for SFR) and $V_\textrm{rot}$ (a proxy for
total mass), as described fully in Bamford et al. (2005b).

Our TFR for 89 field spirals with $0.1 \leq z \leq 1$ is shown in Fig.~4.  The
slope is not significantly different to that measured locally, i.e., no
evidence for mass-dependent evolution.  Examining the TFR residuals versus $z$
(Fig.~2), or fitting the TFR intercept in several redshift bins (Fig.~5), we
find a $B$-band brightening of $1.0\pm0.5$ mag, for a given $V_\textrm{rot}$,
by $z \sim 1$.  Due to selection effects this is an upper limit to the true
evolution, but still implies slower SFR evolution than measured globally.

This suggests that the global SFR evolution of the universe is \emph{not}
driven by the SFR decline in giant spirals.

\begin{figure}
\begin{minipage}{\textwidth}%
\begin{minipage}[t]{0.45\textwidth}%
\includegraphics[width=0.95\textwidth,angle=270]{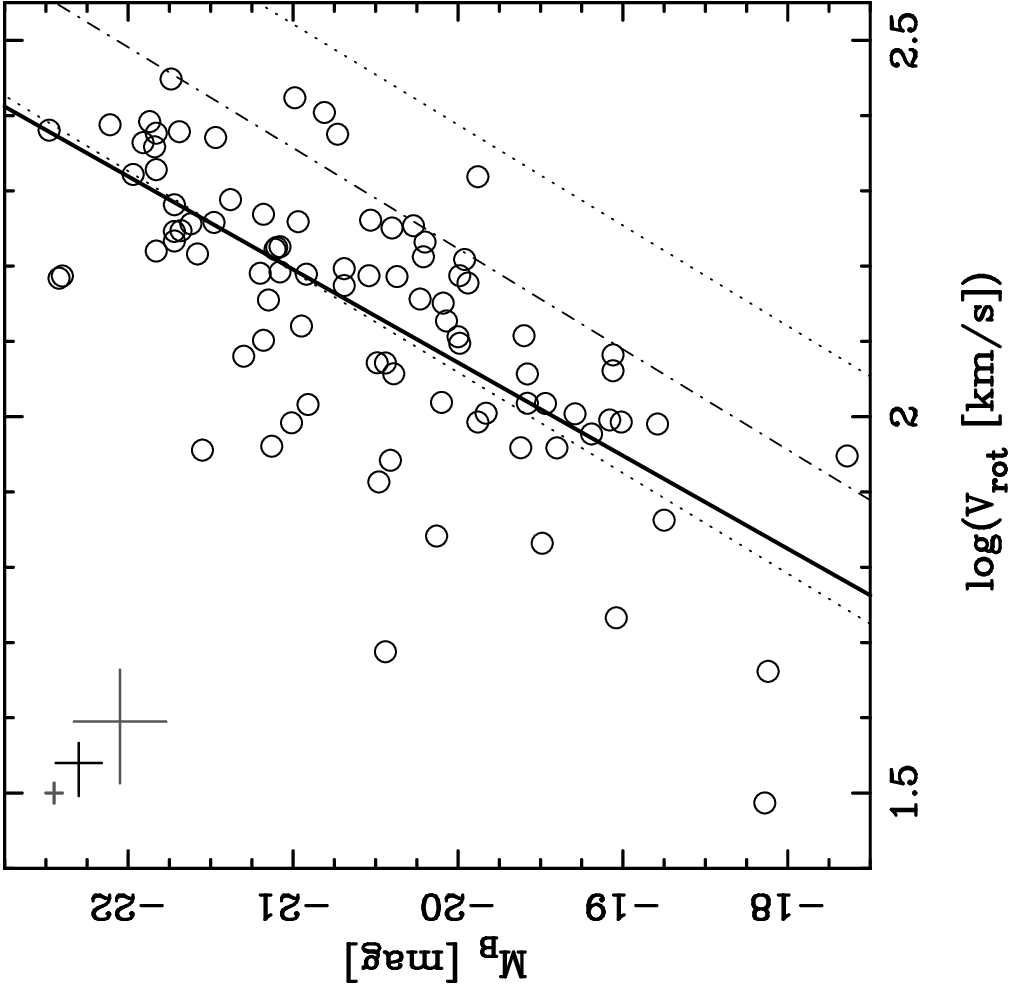}%
\caption{%
  The TFR for our full field TFR sample.  The thin lines indicate the local
  fiducial TFR of PT92, as in Fig.~1.  A fit to the points is
  marked by the thick solid line.  The error bars give the 10th-, 50th-
  and 90th-percentile errors on the points. }%
\vfill
\end{minipage}%
\hfill%
\begin{minipage}[t]{0.45\textwidth}%
\includegraphics[width=0.95\textwidth,angle=270]{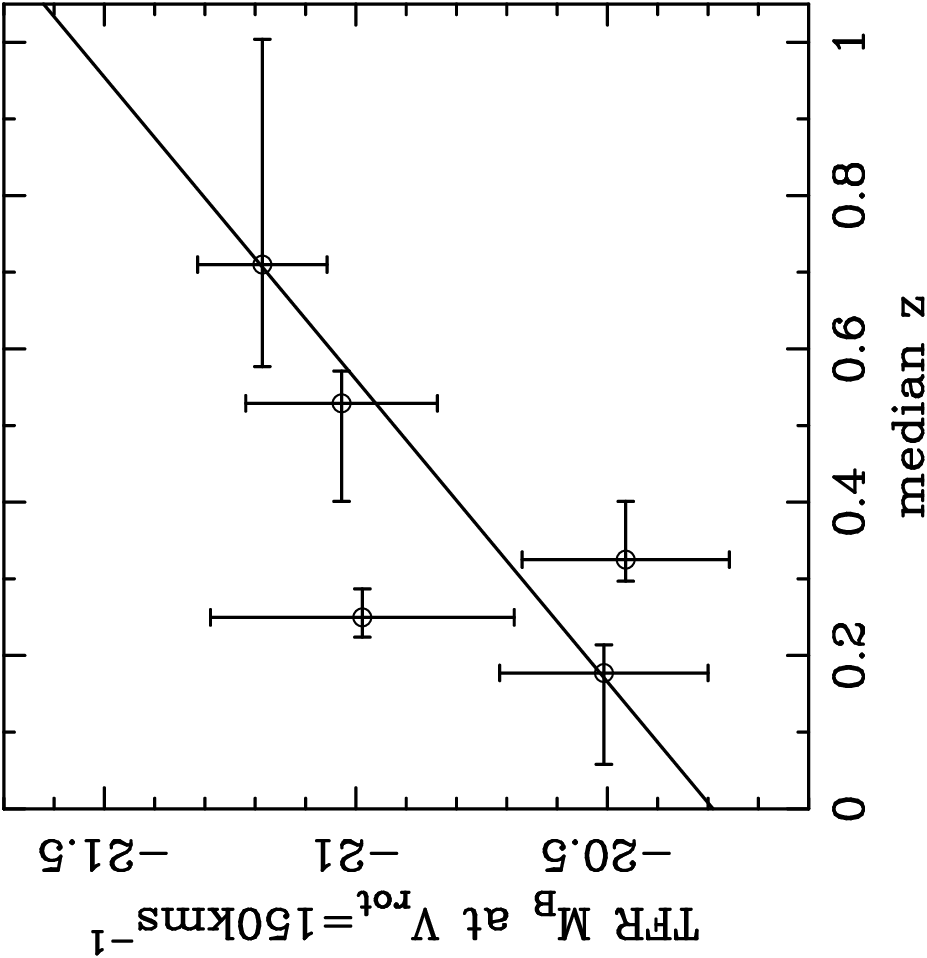}%
\caption{%
  The intercept of TFR fits to subsamples of our field sample in five redshift
  bins.  The fits were performed with constant slope, equal to the PT92 value.
  Horizontal error bars indicate the width of each redshift bin, with the
  point at the median redshift.  Vertical error bars give the uncertainty on
  the fitted intercepts.  The solid line is a fit to the points. }%
\vfill
\end{minipage}%
\end{minipage}%
\end{figure}

\section{Future work with EDisCS}
The ESO Distant Cluster Survey (EDisCS) is a study of galaxies in and around
$\sim\!20$ clusters at $0.4 \leq z_{cl} \leq 1.0$.  These clusters were
optically selected using the Las Campanas Distant Cluster Survey
(\cite{LCDS}), and thus have a wide range of properties.

For each cluster we have a wealth of imaging and spectroscopy data.  Many
projects are underway using these data, including studies of morphology,
luminosity functions, colour-magnitude diagrams, cluster substructure,
velocity distributions, lensing masses and stellar populations.

From the spectra we are now measuring reliable rotation velocities, in order
to perform detailed investigations into the evolution of various galaxy
properties with respect to this useful baseline.  With more than 5 times as
many galaxies, we will have the statistical power to vastly improve upon the
studies shown above.  The addition of HST morphologies, homogeneous colours,
spectral indices and a variety of clusters with well characterised properties,
will allow unprecedented studies of disk galaxy evolution with environment and
redshift.


\begin{chapthebibliography}{1}
\bibitem[{Bamford et~al., 2005a}]{BamfordCluster}  
  {Bamford}, S.P., et al., 2005a, MNRAS, 361, 109
\bibitem[{Bamford et~al., 2005b}]{BamfordField}
  {Bamford}, S.P., {Arag{\' o}n-Salamanca}, A. \& {Milvang-Jensen}, B., 
  2005b, submitted to MNRAS
\bibitem[{Gonzalez et~al., 2001}]{LCDS}
  {Gonzalez}, A.H., et al., 2001, ApJS, 137, 117
\bibitem[{Halliday et~al., 2001}]{Halliday04}
  {Halliday}, C., et al., 2004, A\&A, 427, 397
\bibitem[{Heavens et~al., 2004}]{H04}
  {Heavens}, A., et al., 2004, Nature, 428, 625
\bibitem[{Milvang-Jensen et~al., 2003}]{MJetal03}
  {Milvang-Jensen}, B., et al., 2003, MNRAS, 339, L1
\bibitem[{Pierce \& Tully, 1992}]{PT92}
  {Pierce}, M.J. \&  {Tully}, R.B., 1992, ApJ, 387, 47
\bibitem[{Simard \& Pritchet, 1999}]{SP99}
  {Simard}, L. \& {Pritchet}, C.J., 1999, PASP, 111, 453
\end{chapthebibliography}
\end{document}